# Coupling and decoupling of bilayer graphene monitored by electron energy loss spectroscopy

Yung-Chang Lin[1]*, Amane Motoyama[2], Pablo Solís-Fernández[2], Rika Matsumoto[3], Hiroki Ago[2,4], Kazu Suenaga[1,5]*

[1]Nanomaterials Research Institute, National Institute of Advanced Industrial Science and Technology (AIST), Tsukuba 305-8565, Japan

[2]Interdisciplinary Graduate School of Engineering Sciences, Kyushu University, Fukuoka 816-8580, Japan

[3]Faculty of Engineering, Tokyo Polytechnic University 1583 Iiyama, Atsugi, Kanagawa 243-0297, Japan

[4]Global Innovation Center (GIC), Kyushu University, Fukuoka 816-8580, Japan

[5]The Institute of Scientific and Industrial Research (ISIR-SANKEN), Osaka University, Osaka 567-0047, Japan

**Abstract**

We studied the interlayer coupling and decoupling of bilayer graphene (BLG) by using spatially resolved electron energy loss spectroscopy (EELS) with a monochromated electron source. We correlated the twist angle-dependent energy band hybridization with Moiré superlattices and the corresponding optical absorption peaks. The optical absorption peak originates from the excitonic transition between the hybridized van Hove singularities (vHSs), which shifts systematically with the twist angle. We then proved that the BLG decouples when a monolayer of metal chloride is intercalated in its van der Waals (vdW) gap, and results in the elimination of the vHS peak.

Graphene, the single atomic layer of hexagonal carbon lattice, is the thinnest and strongest, chemically inert, two-dimensional (2D) material[1]. The isolation of single-layer graphene (SLG) has led to the subsequent investigation of various aspects of 2D materials, including growth, characterization, and applications[2]. Furthermore, new types of heteromaterials can be created without being limited by lattice matching and morphology through van der Waals (vdW) integration[3,4]. A number of new materials with unique electronics, optics, and magnetics, are created due to strong interlayer interactions[5–7]. As a result of the development of specimen manipulation and transfer techniques, in addition to stacking different materials, controlling the twist angle of the 2D material stack has been made possible and proved to effectively modulate the electronic and optical properties of the material[8–11]. A recent study reported that stacking of BLG with a specific twist angle will generate flat bands near the Fermi energy to realize pure carbon-based superconductivity at a twist angle of ~1.1°[12].

Conversely, superconductivity has also been realized by intercalating alkali metals between graphite layers for half a century[13–16]. The interlayer coupling and charge transfer between the layers are the major factors that affect the material properties either based on van der Waals stacking or intercalations. The interlayer coupling of BLG is extremely sensitive to the twist angle, and the vHS in the density of states (DOS) has been experimentally measured by scanning tunneling spectroscopy (STS)[17,18]. Also, the excitonic transition between the interlayer vHSs has been studied using optical absorption spectroscopy[19–22]. However, there has been no direct study that explores the interlayer coupling and decoupling of BLG with or without various intercalations. A previous study has suggested the decoupling between the graphite layers after $FeCl_3$ intercalation[23], but there has been no direct evidence of interlayer decoupling that has demonstrated the corresponding optical features together with atomic-level characterization. In addition, it is difficult to determine whether the desired intercalant exists in the vdW gap of BLG or is just adsorbed on the surface by looking at the sample from top view instead of cross-section. Therefore, the ability to simultaneously analyze the atomic structure of an intercalant and know the coupling state between 2D materials is of great technological importance for understanding the properties of intercalation materials.

Herein, we show the coupling and decoupling of BLG upon metal chloride intercalation by using aberration-corrected scanning transmission electron microscopy (STEM) and high-energy resolution electron energy loss spectroscopy (EELS). The twist angle dependent energy band hybridization in coupled BLG reflects to corresponding optical transition peaks. The intercalation induced BLG decoupling is directly detected by monitoring the vanishing of the optical transition peaks by the spatial resolved EELS.

High-quality BLG was grown on a Cu–Ni alloy thin film deposited on sapphire using ambient-pressure chemical vapor deposition[24,25]. The Cu–Ni thin film was etched out in $(NH_4)_2S_2O_8$ solution, and the BLG was transferred to a TEM grid and annealed at 250°C in air for 20 min to remove surface contamination[26]. The BLG/TEM grid was then vacuum-sealed in a Pyrex tube (~$5 \times 10^{-4}$ Pa) together with metal chlorides, that is, $CuCl_2$, $AlCl_3$, and $FeCl_3$ powders[27]. The BLG intercalation was achieved by heating at 150~250°C for 1 h in a tube furnace. The Pyrex tube was then opened in an Ar-filled glove box, and the intercalated BLG/TEM grid was placed on a JEOL vacuum transfer holder and transferred to the TEM chamber without exposing to air. STEM annular dark field (ADF) images were acquired using a JEOL triple-C#3 microscope, an ultra-high vacuum JEM-ARM200F-based microscope equipped with a delta corrector, and a low-voltage cold-field-emission gun. The microscope was operated at 60 kV with a probe current of ~15 pA. The convergence and the inner acquisition semi-angles were set to 37 mrad and 76 mrad, respectively. The ADF images were captured using a Gatan Rio CMOS camera with a resolution of 1024 × 1024 pixels and an acquisition time of 38.5 μs/px. The low-loss EELS spectra were recorded using a JEOL triple-C#2 microscope, a JEM-ARM200F-based microscope equipped with a Schottky field-emission gun, a double Wien filter monochromator, and the delta-type aberration-correctors. The microscope was operated at 60 kV with a probe current of ~8 pA. The convergence semi-angle was 43 mrad, and the inner acquisition semi-angle was 125 mrad. A Gatan Quantum-ERS camera optimized at low voltages was used for high-resolution EELS. The energy resolution was 45 meV after the insertion of 0.5 μm slit. All STEM images and EEL spectra were recorded at room temperature.

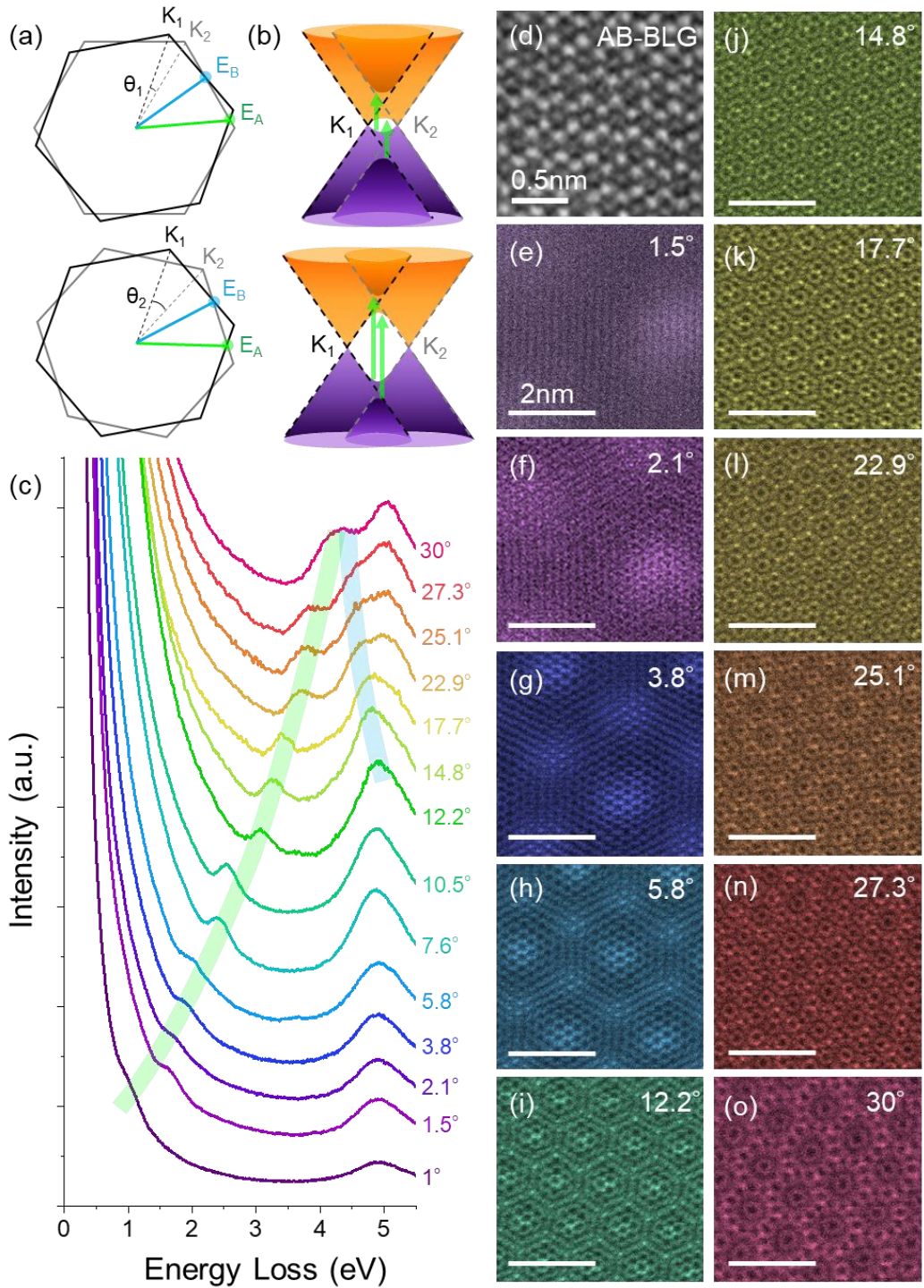

Fig. 1. (a) Schematics of the first Brillouin zone of graphene overlapped with twist angles $\theta_1$ and $\theta_2$ ($\theta_1 < \theta_2$). (b) Schematics of the coupled Dirac cones of twisted-bilayer graphene (BLG) at different twist angles. Optical transitions between the hybridized flat bands in the middle of the two K points are indicated by green arrows. (c) Electron energy loss spectroscopy (EELS) low-loss spectra from BLG with distinct twist angles ($\theta$ = 1–30°). The trends of $E_A$ and $E_B$ van Hove singularity (vHS) transition peaks shifted as a function of the twist angles are highlighted by green and blue ribbons for easier visualization. (d) Annular dark field (ADF) images of AB stacking bilayer graphene. (e-o) ADF images of twisted BLG with twist angles of 1.5°, 2.1°, 3.8°, 5.8°, 12.2°, 14.8°, 17.7°, 22.9°, 25.1°, 27.3°, and 30°, respectively.

First, we show systematic measurements of the "coupling mode" (or vHS transition) from the twisted BLG using the high-energy resolution EELS associated with atomic Moiré imaging. Owing to the linear band structure of graphene, the hybridized BLG band structure can be significantly altered when BLG stacks have a twist angle. The stacking of the two graphene sheets caused the hybridization of their energy bands near the K point. Figure 1a shows the first Brillouin zone of the two graphene layers overlapped with twist angles $\theta_1$ and $\theta_2$ ($\theta_1 < \theta_2$). The two K points ($K_1$ and $K_2$) from different graphene layers are shifted in the momentum space due to the twist angle. The schematics of the energy band structures near the K points of BLG at different twist angles show Dirac cone overlapping and energy-band hybridizations (Fig. 1b). The energy band overlap occurs at the middle, saddle point of $K_1$ and $K_2$ points. The strong interlayer coupling results in energy band hybridization and two split energy flat bands at the saddle point. The flat bands were vHSs with a sharp density of states. Optical transitions ($E_A$ in Fig. 1a) occur from the lower vHS valence band to the lower vHS conduction band, as well as from the higher vHS valence band to the higher vHS conduction band at the saddle point (green arrows in Fig. 1b). Note that the transitions from lower valance band to the higher conduction band, and from higher valance band to lower conduction band at the vHS are optically forbidden transitions[19] which are not observable by EELS. The $E_A$ energy increases with an increase in the twist angle. Another energy band hybridized saddle point occurs near the M point and results in a higher energy optical transition $E_B$, as shown in Fig. 1a and Fig. S1. We systematically correlated the vHS optical transitions with the corresponding BLG twist angles by low-loss EELS with a monochromatic electron source, as shown in Fig. 1c. More than 80 EELS data taken from different twist angles of BLG were analyzed, and 14 spectra were presented with different colors in the θ range from 1° to 30°. The intense zero-loss peaks were not subtracted from the spectra in order to present the original absorption feature at low energy and to avoid the artifact generated through the background subtraction. The corresponding twisted-BLG Moiré superlattices are displayed in Fig. 1e-1o, while an ADF image of AB stacking BLG is shown in Fig. 1d for comparison. The π-plasmon peak of the coupled BLG (4.92 eV) is independent of the twist angle, while there is another distinct peak in each spectrum which appeared in the energy range from 0.5 to 5 eV. This is the optical absorption $E_A$ peak from the vHS excitonic transition, which shifts to higher energies as the twist angle increases (see green ribbon for eye guidance in Fig. 1c). Conversely, the $E_B$ peak caused by the coupling of the high-energy band (Fig. S1) almost overlaps with the π-plasmon peak at lower twist angles and results in slightly red shift of the π-plasmon peak. It gradually shifts to a lower energy at higher twist angles and finally overlaps with $E_A$ at a twist angle of 30° (see blue ribbon for eye guidance in Fig. 1c). These vHS peaks are unique optical signatures of the twisted BLG. Our experimental data are consistent with previously reported tight-binding models[19].

The interlayer hybridization resulting in flat bands should have localized wave function profiles in real space especially for a twisted BLG of small angle. It should show local density of states highly concentrated in the moiré AA stacked region but depleted in the AB stacked region[28–31]. Such kind of spatial electron density variation and the corresponding band structure differences are significant only for the twist angle close to 0°. In our experiment, the electron source possesses zero momentum (q=0) which excites the dipole transition in materials. More importantly, the delocalization of dipole transition in the current energy-loss range is estimated as large as 5-10 nm, therefore the moiré-induced bandgap variation is indeed smeared out in our experiment at the optical limit (q=0) (Fig. S2). Further experiment with dark-field EELS (q>0) would be more beneficial to reduce the EELS delocalization and to perceive the moiré modulations, but this is not within the scope of the paper. The EELS electronic excitations in valence range (below 5 eV) should involve many-body physics, with the spectral features renowned for rich single-particle, collective, and mixed characters. Here we concentrate on the single-particle optical absorption due to vHS, and the arguments in the other related physics, such as plasmonic (dielectric or collective) response, will not be discussed here.

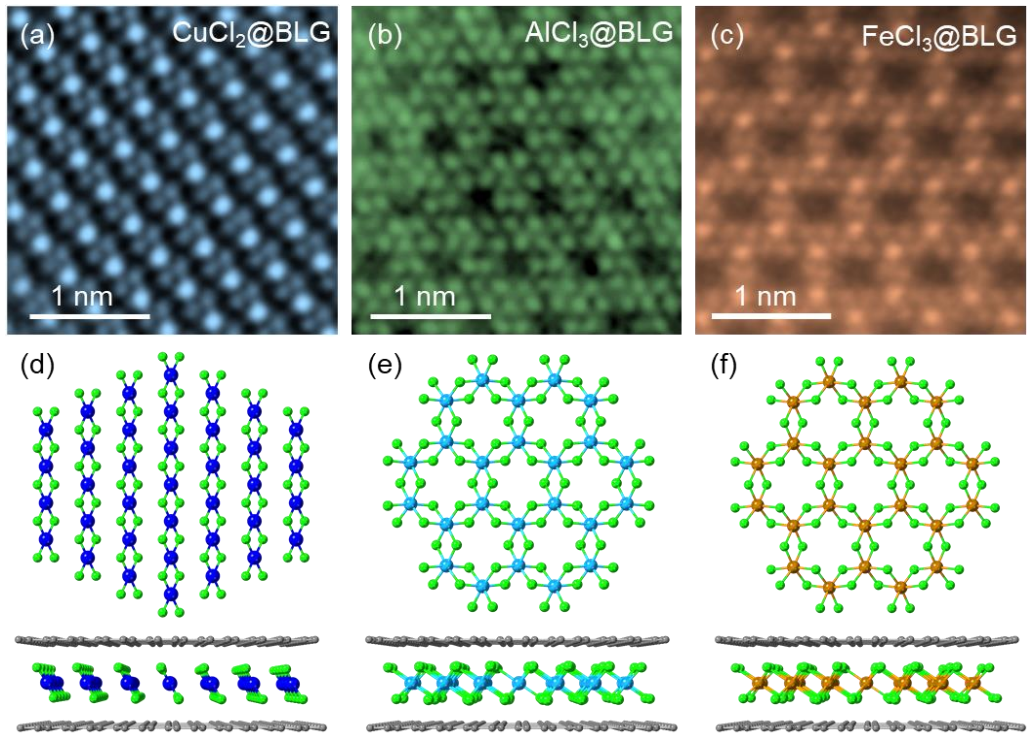

Fig. 2. (a-c) ADF images of CuCl$_2$@BLG, AlCl$_3$@BLG, and FeCl$_3$@BLG. (d-f) Corresponding atomic models from top and side views.

We show the disappearance of this "coupling mode" (or vHS transition of BLG) upon intercalation. Figure 2 summarizes a series of metal chlorides intercalated in BLGs and their distinct atomic structures. Figure 2a shows an ADF image of a monolayer CuCl$_2$ intercalated in the BLG gap (CuCl$_2$@BLG). The intercalated CuCl$_2$ possesses trigonal symmetry and consists of long parallel chains, where the spacing between two parallel CuCl$_2$ chains is ~3.6 Å. The corresponding atomic model shown in Fig. 2d suggests that CuCl$_2$ has a structure distinct from that of bulk octahedral (1T) MCl$_2$. Figures 2b and 2c display the ADF images of AlCl$_3$@BLG and FeCl$_3$@BLG, respectively. These two intercalation compounds are both octahedral structures consisting of one-third of vacancies at the metal sites, as shown by the atomic models in Figs. 2e and 2f. AlCl$_3$ and FeCl$_3$ can be distinguished from the ADF intensity of the metal site, which is proportional to the atomic number (C:6, Al:13, Cl:17, Fe:26, Cu:29)[32]. BLG lattice contrast is barely observed in the metal chloride-intercalated area.

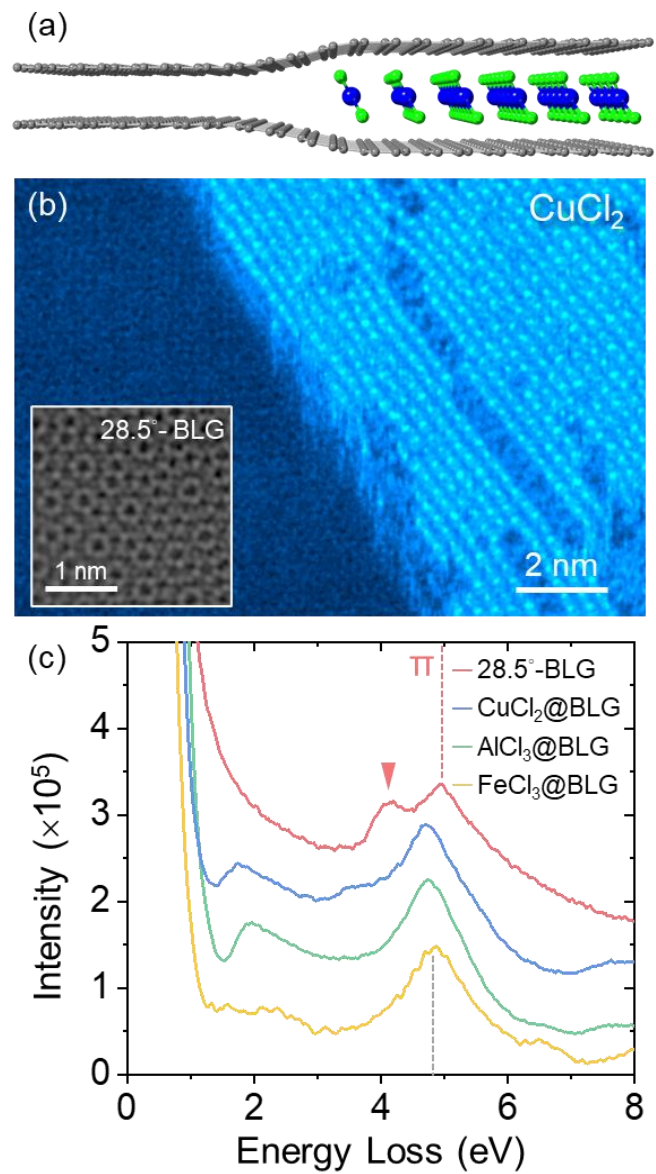

Fig. 3. (a) Schematic of a cross-sectional view of BLG with an intercalant (CuCl$_2$). (b) An ADF image of an edge of CuCl$_2$@BLG. The inset is a filtered ADF image of the graphene domain. (c) EELS low-loss profiles of 28.5° twisted BLG, $CuCl_2$@BLG, $AlCl_3$@BLG, and $FeCl_3$@BLG. The arrow at ~4 eV indicates the vHS transition ("coupling mode") of empty BLG, which is absent in the intercalated BLGs. The twist angle of BLG for $AlCl_3$ intercalation is 25.6° and is 20.9° for $FeCl_3$ intercalation.

Figure 3 shows a spatially resolved EELS study across an intercalation frontier. The schematic in Fig. 3a shows a cross-sectional view of the investigated structure. The interlayer distance between two graphene layers of empty BLG is ~3.4 Å on the left side and is expanded along the c-axis by the intercalation at the right side. Figure 3b shows an ADF image of half-filled $CuCl_2$@BLG with clearly resolved lattices of both the BLG (left) and $CuCl_2$-intercalated domains (right). The BLG is stacked with a twist angle of 28.5°, and results in a unique Moiré pattern, as shown in the inset of Fig. 3b. An EELS line scan was performed by scanning the probe from the BLG domain to the $CuCl_2$ intercalated domain (Fig. S3). The valence electron excitation in the low-energy (optical) range (0.5~10 eV) in EELS is an analogue of optical and infrared absorption spectroscopy[33,34]. The EEL low-loss spectra of the twisted BLG (28.5°) and $CuCl_2$@BLG are displayed in the red and blue profiles in Fig. 3c. In the BLG spectrum, a π-plasmon peak appears at 4.92 eV, and a sharp peak appears at 4 eV (pointed by the red triangle); this is the specific optical signature of the twisted BLG (at 28.5°) originated from the interlayer transition between the vHSs. It is important that the vHS peak ("coupling mode") vanishes in the $CuCl_2$@BLG spectrum. This indicates the decoupling of the two graphene layers. Effective blocking of the electron coupling between the two graphene layers by a middle layer has never been reported experimentally nor theoretically. The π-plasmon peak is slightly red-shifted to 4.68 eV, reflecting the optical transition of SLG[35]. The linear band structure of SLG results in a smooth absorption profile and shows a lower absorption background than that of BLG in the low-energy range (< 3 eV); therefore, the spectrum of CuCl$_2$@BLG mainly reflects the optical features of the intercalants.

We then generalized the decoupling of BLG for other metal chloride intercalations. The green and yellow profiles are the absorption spectra of AlCl$_3$@BLG and FeCl$_3$@BLG, respectively. The surface deposited and the intercalated metal chlorides are difficult to be distinguished by ADF contrast only, but these two configurations behave differently in stability and in spectroscopy. The intercalated metal chlorides are much stable under e-beam scanning and can even undergo phase transformation between BLG[36], on the other hand, the surface deposited ones are easily removed by e-beam (Fig. S4 and Movie 1). The absence of vHS peak ("coupling mode") is another proof that the metal chloride is indeed intercalated in the BLG gap and that it does not exist on the surface, thus indicating the decoupling of the two graphene layers due to the existence of intercalants in between the gaps. The vHS peak remains for the coupled BLG with surface deposited metal chlorides (Fig. S5a). Besides, the surface deposited metal chlorides are mostly oxidized in comparison with the intercalated ones (Fig. S5b). With this information we conclude the metal chlorides presented in this paper are indeed intercalated.

In general, graphene is a useful 2D support film for the investigation of the bandgap energy and the related optrtical absorption properties of nanomaterials below the graphene plasmon (4~5 eV) using atomic resolution electron probes in STEM[37]. Therefore, these profiles can be regarded as the excitonic fingerprints of monolayer CuCl$_2$, AlCl$_3$, and FeCl$_3$ supported with graphene. The detailed interpretation of spectral features for the intercalated BLGs requires a full theoretical study involving both the excitonic effect and dielectric response, and then is not in the scope of this paper.

In low-dimensional materials, the surface state and the interaction between layers are the key factors affecting the material properties. The realization of atomic-level analysis of the structure of nanomaterials and the energy band changes caused by surface interactions is necessary for future advances in nanotechnology. Using state-of-the-art electron microscopy and spectroscopy techniques, we successfully established the relationship between the twist angle, the Moiré superlattices, and the absorption spectrum caused by the interlayer coupling of BLG. This technique can also be used for other layered integrated nanomaterials to understand the coupling conditions between layers of the same or different materials, as well as the energy band variations in localized regions associated with Moiré interactions.

**Supporting Information**

The supporting Information is available free of charge on the ACS Publications website https://jpn01.safelinks.protection.outlook.com/?url=http%3A%2F%2Fpubs.acs.org%2F&data=04%7C01%7Cyc-lin%40aist.go.jp%7Ca52491e690a74027edfe08d9b375af0d%7C18a7fec8652f409b8369272d9ce80620%7C0%7C0%7C637738140295580980%7CUnknown%7CTWFpbGZsb3d8eyJWIjoiMC4wLjAwMDAiLCJQIjoiV2luMzIiLCJBTiI6Ik1haWwiLCJXVCI6Mn0%3D%7C3000&sdata=yaVspXs%2BH3bbHhI3szV6x8aTVBj6QsAiuIwkCBYFOmQ%3D&reserved=0.

Schematic description of the optical transitions $E_A$ and $E_B$. Discussion of EELS spatial resolution and delocalization at the optical limit (q=0). Additional experimental proof of the vHS transition peak vanishes at the metal chloride ($CuCl_2$) intercalated BLG domain. Stability difference between the surface deposited and the intercalated metal chloride ($CuCl_2$) under e-beam. EELS characterization of the surface deposited and the intercalated metal chloride ($AlCl_3$).

**Acknowledgements**

Y.-C.L. and K.S. acknowledge JSPS-KAKENHI [(16H06333), (18K14119)], the JST-CREST program (JPMJCR20B1, JMJCR20B5, JPMJCR1993), the JSPS A3 Foresight Program, and the Kazato Research Encouragement Prize. HA acknowledges JSPS-KAKENHI (18H03864, 19K22113, 21K18878) and the JST-CREST program (JPMJCR20B1).

**References**

(1) Geim, A. K.; Novoselov, K. S. The Rise of Graphene. *Nat. Mater.* **2007**, *6*, 183–191.

(2) Chhowalla, M.; Shin, H. S.; Eda, G.; Li, L.-J.; Loh, K. P.; Zhang, H. The Chemistry of Two-Dimensional Layered Transition Metal Dichalcogenide Nanosheets. *Nat. Chem.* **2013**, *5*, 263–275.

(3) Novoselov, K. S.; Mishchenko, A.; Carvalho, A.; Castro Neto, A. H. 2D Materials and van Der Waals Heterostructures. *Science* **2016**, *353*, acc9439.

(4) Liu, Y.; Huang, Y.; Duan, X. Van Der Waals Integration before and beyond Two-Dimensional Materials. *Nature* **2019**, *567*, 323–333.

(5) Jariwala, D.; Marks, T. J.; Hersam, M. C. Mixed-Dimensional van Der Waals

Heterostructures. *Nat. Mater.* **2017**, *16*, 170–181.

(6) Li, B.; Wan, Z.; Wang, C.; Chen, P.; Huang, B.; Cheng, X.; Qian, Q.; Li, J.; Zhang, Z.; Sun, G.; Zhao, B.; Ma, H.; Wu, R.; Wei, Z.; Liu, Y.; Liao, L.; Ye, Y.; Huang, Y.; Xu, X.; Duan, X.; Ji, W.; Duan, X. Van Der Waals Epitaxial Growth of Air-Stable CrSe2 Nanosheets with Thickness-Tunable Magnetic Order. *Nat. Mater.* **2021**, *20*, 818–825.

(7) Zhao, X.; Song, P.; Wang, C.; Riis-Jensen, A. C.; Fu, W.; Deng, Y.; Wan, D.; Kang, L.; Ning, S.; Dan, J.; Venkatesan, T.; Liu, Z.; Zhou, W.; Thygesen, K. S.; Luo, X.; Pennycook, S. J.; Loh, K. P. Engineering Covalently Bonded 2D Layered Materials by Self-Intercalation. *Nature* **2020**, *581*, 171–177.

(8) Yoo, H.; Engelke, R.; Carr, S.; Fang, S.; Zhang, K.; Cazeaux, P.; Sung, S. H.; Hovden, R.; Tsen, A. W.; Taniguchi, T.; Watanabe, K.; Yi, G. C.; Kim, M.; Luskin, M.; Tadmor, E. B.; Kaxiras, E.; Kim, P. Atomic and Electronic Reconstruction at the van Der Waals Interface in Twisted Bilayer Graphene. *Nature Materials*. 2019, pp 448–453.

(9) Wang, L.; Shih, E. M.; Ghiotto, A.; Xian, L.; Rhodes, D. A.; Tan, C.; Claassen, M.; Kennes, D. M.; Bai, Y.; Kim, B.; Watanabe, K.; Taniguchi, T.; Zhu, X.; Hone, J.; Rubio, A.; Pasupathy, A. N.; Dean, C. R. Correlated Electronic Phases in Twisted Bilayer Transition Metal Dichalcogenides. *Nat. Mater.* **2020**, *19*, 861–866.

(10) Liu, E.; Barré, E.; van Baren, J.; Wilson, M.; Taniguchi, T.; Watanabe, K.; Cui, Y. T.; Gabor, N. M.; Heinz, T. F.; Chang, Y. C.; Lui, C. H. Signatures of Moiré Trions in WSe$_2$/MoSe$_2$ Heterobilayers. *Nature* **2021**, *594*, 46–50.

(11) Alexeev, E. M.; Ruiz-tijerina, D. A.; Danovich, M.; Hamer, M. J.; Terry, D. J.;

Nayak, P. K.; Ahn, S.; Pak, S.; Lee, J.; Sohn, J. I.; Molas, M. R.; Koperski, M.; Watanabe, K.; Shin, H. S.; Fal, V. I.; Tartakovskii, A. I. Resonantly Hybridized Excitons in Moiré Superlattices in van Der Waals Heterostructures. *Nature*. 2019, p 81.

(12) Cao, Y.; Fatemi, V.; Fang, S.; Watanabe, K.; Taniguchi, T.; Kaxiras, E.; Jarillo-Herrero, P. Unconventional Superconductivity in Magic-Angle Graphene Superlattices. *Nature* **2018**, *556*, 43–50.

(13) Hannay, N. B.; Geballe, T. H.; Matthias, B. T.; Andres, K.; Schmidt, P.; MacNair, D. Superconductivity in Graphitic Compounds. *Phys. Rev. Lett.* **1965**, *14*, 225–226.

(14) Emery, N.; Hérold, C.; D'Astuto, M.; Garcia, V.; Bellin, C.; Marêché, J. F.; Lagrange, P.; Loupias, G. Superconductivity of Bulk $CaC_6$. *Phys. Rev. Lett.* **2005**, *95*, 087003.

(15) Weller, T. E.; Ellerby, M.; Saxena, S. S.; Smith, R. P.; Skipper, N. T. Superconductivity in the Intercalated Graphite Compounds C6Yb and C6Ca. *Nat. Phys.* **2005**, *1*, 39–41.

(16) Kanetani, K.; Sugawara, K.; Sato, T.; Shimizu, R.; Iwaya, K.; Hitosugi, T.; Takahashi, T. Ca Intercalated Bilayer Graphene as a Thinnest Limit of Superconducting $C_6$Ca. *Proc. Natl. Acad. Sci. U. S. A.* **2012**, *109*, 19610–19613.

(17) Li, G.; Luican, A.; Lopes Dos Santos, J. M. B.; Castro Neto, A. H.; Reina, A.; Kong, J.; Andrei, E. Y. Observation of Van Hove Singularities in Twisted Graphene Layers. *Nat. Phys.* **2010**, *6*, 109–113.

(18) Brihuega, I.; Mallet, P.; González-Herrero, H.; Trambly De Laissardière, G.; Ugeda, M. M.; Magaud, L.; Gómez-Rodríguez, J. M.; Ynduráin, F.; Veuillen, J.

Y. Unraveling the Intrinsic and Robust Nature of van Hove Singularities in Twisted Bilayer Graphene by Scanning Tunneling Microscopy and Theoretical Analysis. *Phys. Rev. Lett.* **2012**, *109*, 196802.

(19) Moon, P.; Koshino, M. Optical Absorption in Twisted Bilayer Graphene. *Phys. Rev. B* **2013**, *87*, 205404.

(20) Basile, L. A.; Zhou, W.; Salafranca, J.; Idrobo, J.-C. Direct Observation of the Optical Response of Twisted Bilayer Graphene by Electron Energy Loss Spectroscopy. *Microsc. Microanal.* **2013**, *19*, 1920–1921.

(21) Zhou, W.; Dellby, N.; Basile, L.; Aoki, T.; Salafranca, J.; Mardinly, J.; Carpenter, R.; Krivanek, O. L.; Idrobo, J. C.; Pennycook, S. J. Monochromatic STEM-EELS for Correlating the Atomic Structure and Optical Properties of Two-Dimensional Materials. *Microsc. Microanal.* **2014**, *20*, 96–97.

(22) Havener, R. W.; Liang, Y.; Brown, L.; Yang, L.; Park, J. Van Hove Singularities and Excitonic Effects in the Optical Conductivity of Twisted Bilayer Graphene. *Nano Lett.* **2014**, *14*, 3353–3357.

(23) Zhao, W.; Tan, P. H.; Liu, J.; Ferrari, A. C. Intercalation of Few-Layer Graphite Flakes with $FeCl_3$: Raman Determination of Fermi Level, Layer by Layer Decoupling, and Stability. *J. Am. Chem. Soc.* **2011**, *133*, 5941–5946.

(24) Takesaki, Y.; Kawahara, K.; Hibino, H.; Okada, S.; Tsuji, M.; Ago, H. Highly Uniform Bilayer Graphene on Epitaxial Cu-Ni(111) Alloy. *Chem. Mater.* **2016**, *28*, 4583–4592.

(25) Solís-Fernández, P.; Terao, Y.; Kawahara, K.; Nishiyama, W.; Uwanno, T.; Lin, Y.-C.; Yamamoto, K.; Nakashima, H.; Nagashio, K.; Hibino, H.; Suenaga, K.; Ago, H. Isothermal Growth and Stacking Evolution in Highly Uniform Bernal-

Stacked Bilayer Graphene. *ACS Nano* **2020**, *14*, 6834–6844.

(26) Lin, Y. C.; Lu, C. C.; Yeh, C. H.; Jin, C.; Suenaga, K.; Chiu, P. W. Graphene Annealing: How Clean Can It Be? *Nano Lett.* **2012**, *12*, 414–419.

(27) Kinoshita, H.; Jeon, I.; Maruyama, M.; Kawahara, K.; Terao, Y.; Ding, D.; Matsumoto, R.; Matsuo, Y.; Okada, S.; Ago, H. Highly Conductive and Transparent Large-Area Bilayer Graphene Realized by $MoCl_5$ Intercalation. *Adv. Mater.* **2017**, *29*, 1702141.

(28) Trambly De Laissardière, G.; Mayou, D.; Magaud, L. Numerical Studies of Confined States in Rotated Bilayers of Graphene. *Phys. Rev. B - Condens. Matter Mater. Phys.* **2012**, *86*, 125413.

(29) Lopes Dos Santos, J. M. B.; Peres, N. M. R.; Castro Neto, A. H. Continuum Model of the Twisted Graphene Bilayer. *Phys. Rev. B - Condens. Matter Mater. Phys.* **2012**, *86*, 155449.

(30) Bistritzer, R.; MacDonald, A. H. Moiré Bands in Twisted Double-Layer Graphene. *Proc. Natl. Acad. Sci. U. S. A.* **2011**, *108*, 12233–12237.

(31) Shi, H.; Zhan, Z.; Qi, Z.; Huang, K.; Veen, E. van; Silva-Guillén, J. Á.; Zhang, R.; Li, P.; Xie, K.; Ji, H.; Katsnelson, M. I.; Yuan, S.; Qin, S.; Zhang, Z. Large-Area, Periodic, and Tunable Intrinsic Pseudo-Magnetic Fields in Low-Angle Twisted Bilayer Graphene. *Nat. Commun.* **2020**, *11*, 371.

(32) Krivanek, O. L.; Chisholm, M. F.; Nicolosi, V.; Pennycook, T. J.; Corbin, G. J.; Dellby, N.; Murfitt, M. F.; Own, C. S.; Szilagyi, Z. S.; Oxley, M. P.; Pantelides, S. T.; Pennycook, S. J. Atom-by-Atom Structural and Chemical Analysis by Annular Dark-Field Electron Microscopy. *Nature* **2010**, *464*, 571–574.

(33) Senga, R.; Pichler, T.; Suenaga, K. Electron Spectroscopy of Single Quantum

Objects to Directly Correlate the Local Structure to Their Electronic Transport and Optical Properties. *Nano Lett.* **2016**, *16*, 3661–3667.

(34) Tizei, L. H. G.; Lin, Y.-C.; Mukai, M.; Sawada, H.; Lu, A.-Y.; Li, L.-J.; Kimoto, K.; Suenaga, K. Exciton Mapping at Subwavelength Scales in Two-Dimensional Materials. *Phys. Rev. Lett.* **2015**, *114*, 107601.

(35) Wachsmuth, P.; Hambach, R.; Benner, G.; Kaiser, U. Plasmon Bands in Multilayer Graphene. *Phys. Rev. B* **2014**, *90*, 235434.

(36) Lin, Y. C.; Motoyama, A.; Kretschmer, S.; Ghaderzadeh, S.; Ghorbani-Asl, M.; Araki, Y.; Krasheninnikov, A. V.; Ago, H.; Suenaga, K. Polymorphic Phases of Metal Chlorides in the Confined 2D Space of Bilayer Graphene. *Adv. Mater.* **2021**, *2105898*, 1–8.

(37) Lin, J.; Gomez, L.; De Weerd, C.; Fujiwara, Y.; Gregorkiewicz, T.; Suenaga, K. Direct Observation of Band Structure Modifications in Nanocrystals of $CsPbBr_3$ Perovskite. *Nano Lett.* **2016**, *16*, 7198–7202.

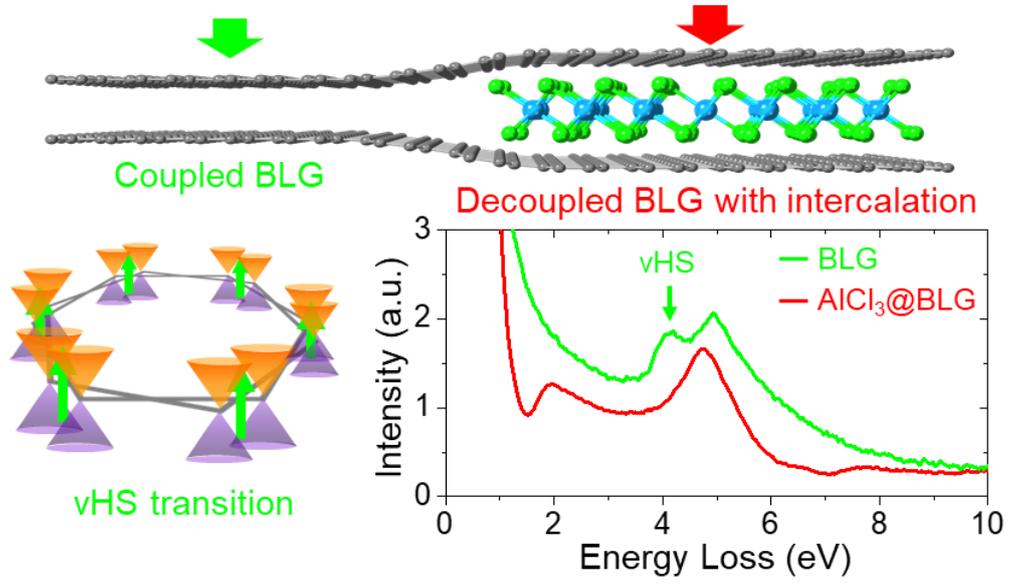

TOC graphic